\documentclass[11pt]{amsart}
\usepackage{amsthm}
\usepackage{amsmath,amstext,amsthm,amsfonts,amssymb,amsthm}
\usepackage{multicol}
\usepackage{latexsym}
\usepackage{color}
\usepackage{graphicx}
\usepackage{epsf}
\usepackage{epsfig}
\usepackage{epic}
\usepackage{graphics}
\usepackage{verbatim}
\usepackage{color}
\usepackage{hyperref}
\newtheorem{theorem}{Theorem}[section]

\theoremstyle{definition}
\newtheorem{definition}[theorem]{Definition}
\newtheorem{proposition}[theorem]{Proposition}
\newtheorem{example}[theorem]{Example}

\theoremstyle{remark}
\newtheorem{remark}[theorem]{Remark}
\numberwithin{equation}{section}

\newcommand{\N}{\mathcal{N}}

\newcommand{\HI}{\mathfrak{H}}

\newcommand{\C}{\mathbb{C}}

\newcommand{\oz}{\overline{z}}
\newcommand{\qu}{\mathbf{q}}
\newcommand{\pu}{\mathbf{p}}
\newcommand{\oqu}{\overline{\mathbf{q}}}

\begin{document}
\title[Quantization of quaternions ]{Coherent state Quantization of quaternions}
\author{B. Muraleetharan$^{\dagger}$, K. Thirulogasanthar$^{\ddagger}$}
\address{$^{\dagger}$ Department of mathematics and Statistics, University of Jaffna, Thirunelveli, Sri Lanka.}
\address{$^{\ddagger}$ Department of Computer Science and Software Engineering, Concordia University, 1455 De Maisonneuve Blvd. West, Montreal, Quebec, H3G 1M8, Canada.}
\email{santhar@gmail.com and bbmuraleetharan@jfn.ac.lk}
\date{\today}
%\dedicatory{This paper is dedicated to our authors.}
%%%%%%%%%%%%%%%%%%%%%%%%%%%%%%%%%%%%%%%%%%%%%%%%%%%%%%%%%%%%%%%%%%%%%%%%%%%%%%%%%%%%%%%%%%%%%%%%%%%%%%%%%%%%%%%%%%%%%%%%%%%%%%%%%%%%%%%%%%%%%%%%%%%%%%%%%%%%%%%%%%%%%%%
%%%%%%%%%%%%%%%%%%%%%%%%%%%%%%%%%%%%%%%%
%\maketitle
\pagestyle{myheadings}
%%%%%%%%%%%%%%%%%%%%%%%%%%%%%%%%%%%%%%%%%%%%%%%%%%%%%%%%%%%%%
%\begin{multicols}{2}
\begin{abstract}
Parallel to the quantization of the complex plane, using the canonical coherent states of a right quaternionic Hilbert space, quaternion field of quaternionic quantum mechanics is quantized.  Associated upper symbols, lower symbols and related quantities are analyzed.  Quaternionic version of the harmonic oscillator and Weyl-Heisenberg algebra are also obtained.
\end{abstract}
\maketitle
%%%%%%%%%%%%%%%%%%%%%%%%%%%%%%%%%%%%%%%%%%%%%%%%%%%%%%%%%%%%%%%%%%%%%%%%%%%%%%%%%%%%%%%%%%%%%%%%%%%%%%%%%%%%%%%%%%%%%%%%%%%%%%
\section{Introduction}\label{sec_intro}
Quantization is commonly understood as the transition from classical to quantum mechanics. One may also say, to a certain extent, quantization relates to a larger discipline than just restricting to specific domains of physics.  In physics, quantization is a procedure that associates with an algebra $A_{cl}$ of classical observables an algebra $A_q$ of quantum observables. The algebra $A_{cl}$ is usually realized as a commutative Poisson algebra of derivable functions on a symplectic (or phase) space $X$. The algebra $A_q$ is, however, non-commutative in general and the quantization procedure must provide a correspondence $A_{cl}\mapsto A_q~:~f\mapsto A_f$. Most physical quantum theories may be obtained as the result of a canonical quantization procedure. However, among the various quantization procedures available in the literature, the coherent state quantization (CS quantization) appear quite arbitrary because the only structure that a space $X$ must possess is a measure. Once a family of CS or frame labeled by a measure space $X$ is given one can quantize the measure space $X$. Various quantization schemes and their advantages and drawbacks are discussed in detail, for example, in \cite{AE,Gaz, Per, Ali}.\\

Due to the non commutativity of quaternions, quaternionic Hilbert spaces are formed by right or left multiplication of vectors by quaternionic scalars; the two different conventions give isomorphic versions of the theory. Quaternions can always be represented, through symplectic component functions, as a pair of complex numbers and thereby quaternions possess a symplectic structure. However,  quaternionic quantum mechanics is inequivalent to complex quantum mechanics.  In analogy with complex quantum mechanics, states of quaternionic quantum mechanics are described by vectors of a separable quaternionic Hilbert space and observables in quaternionic quantum mechanics are represented by quaternion linear and self-adjoint operators \cite{Ad}.\\

The CS quantization in the complex quantum mechanics is a well-known and well-studied problem. Using the method of CS quantization, various phase spaces such as complex field, complex unit disc, circle in complex plane, and cylindrical phase spaces , to name a few, have been quantized \cite{Ali, AGH, Gaz, Ga, Nic}. However, quantization of the quaternion field has not been studied yet. In this regard, parallel to the (CS) quantization of the complex field, in this note, we present CS quantization of the quaternion field using CS of a right quaternionic Hilbert space, compute upper and lower symbols, and study the matrix elements. The quaternionic version of the harmonic oscillator and Weyl-Heisenberg algebra are also obtained. Since the properties of operators from complex Hilbert spaces do not directly translate to the operators on quaternionic Hilbert spaces \cite{Gi}, we shall investigate quaternionic operator properties associated with the quantization as needed.\\

The rest of the paper is organized as follows. In section 2 we present mathematical preliminaries as required for the development of the article. In fact section 2 deals with some properties of quaternion, a measure on quaternion, definition of quaternion slices and quaternion Hilbert spaces.  CS on a right quaternion Hilbert space is presented in section 3. Section 4 deals with the well-known general scheme of CS quantization. Using the CS developed in section 3 we CS quantize the field of quaternions in section 5 where we also study the associated upper symbols, lower symbols, a Hamiltonian structure and oscillator algebra. Further, we realize the upper symbols as differential operators in terms of the so-called Cullen derivatives. In section 6, as illustrative examples, we present quantization map with one and two indexed quaternionic Hermite polynomials. Section 7 ends the manuscript with a conclusion.

%%%%%%%%%%%%%%%%%%%%%%%%%%%%%%%%%%%%%%%%%%%%%%%%%%%%%%%%%%%%%%%%%%%%%%%%%%%%%%%%%%%%%%%%%%%%%%%%%%%%%%%%%%%%%%%%%%%%%%%%%%%%%%%%%%%%%%%%%%%%%%%%%%%%%%%%%%%%%%%%%%%%%%%
\section{Mathematical preliminaries}
In order to make the paper self-contained, we recall few facts about quaternions which may not be well-known. In particular, we revisit the $2\times 2$ complex matrix representations of quaternions, quaternionic Hilbert spaces as needed here. For details we refer the reader to \cite{Ad, Gi, Vis, Za}.
\subsection{Quaternions}
Let $H$ denote the field of quaternions. Its elements are of the form $\qu=x_0+x_1i+x_2j+x_3k$ where $x_0,x_1,x_2$ and $x_3$ are real numbers, and $i,j,k$ are imaginary units such that $i^2=j^2=k^2=-1$, $ij=-ji=k$, $jk=-kj=i$ and $ki=-ik=j$. The quaternionic conjugate of $\qu$ is defined to be $\overline{\qu} = x_0 - x_1i - x_2j - x_3k$. We shall find it convenient to use the representation of quaternions by $2\times 2$ complex matrices:
 \begin{equation}
\qu = x_0 \sigma_{0} + i \underline{x} \cdot \underline{\sigma},
 \end{equation}
with $x_0 \in \mathbb R , \quad \underline{x} = (x_1, x_2, x_3)
\in \mathbb R^3$, $\sigma_0 = \mathbb{I}_2$, the $2\times 2$ identity matrix, and
$\underline{\sigma} = (\sigma_1, -\sigma_2, \sigma_3)$, where the
$\sigma_\ell, \; \ell =1,2,3$ are the usual Pauli matrices. The quaternionic imaginary units are identified as, $i = \sqrt{-1}\sigma_1, \;\; j = -\sqrt{-1}\sigma_2, \;\; k = \sqrt{-1}\sigma_3$. Thus,
 \begin{equation}
\qu = \left(\begin{array}{cc}
x_0 + i x_3 & -x_2 + i x_1 \\
x_2 + i x_1 & x_0 - i x_3
\end{array}\right) \qquad %\text{and} \qquad \overline{\qu} = \qu^\dag\quad \text{(matrix adjoint)}\; .
 \label{q3}
 \end{equation}
and $\overline{\qu} = \qu^\dag\quad \text{(matrix adjoint)}\; .$
Introducing the polar coordinates:
 \begin{eqnarray*}
x_0 &=& r \cos{\theta}, \\
x_1 &=& r \sin{\theta} \sin{\phi} \cos{\psi}, \\
x_2 &=& r \sin{\theta} \sin{\phi} \sin{\psi}, \\
x_3 &=& r \sin{\theta} \cos{\phi},
 \end{eqnarray*}
where $(r,\phi,\theta,\psi)  \in [0,\infty)\times[0,\pi]\times[0,2\pi)^{2}$, we may write
 \begin{equation}
\qu = A(r) e^{i \theta \sigma(\widehat{n})}
 \label{q4},
 \end{equation}
where
 \begin{equation}
A(r) = r\mathbb \sigma_0 
 \end{equation}
and\begin{equation}
\sigma(\widehat{n}) = \left(\begin{array}{cc}
\cos{\phi} & \sin{\phi} e^{i\psi} \\
\sin{\phi} e^{-i\psi} & -\cos{\phi}
\end{array}\right).
\label{q5}\end{equation}
The matrices
$A(r)$ and $\sigma(\widehat{n})$ satisfy the conditions,
 \begin{equation}
A(r) = A(r)^\dagger,~\sigma(\widehat{n})^2 = \sigma_0
,~\sigma(\widehat{n})^\dagger = \sigma(\widehat{n})
 \label{san1}
 \end{equation}
and
%\begin{equation}
$\lbrack A(r), \sigma(\widehat{n}) \rbrack = 0.$
 %\label{san1}
 %\end{equation}
Note that a real norm on $H$ is defined by  $$\vert\qu\vert^2  := \overline{\qu} \qu = r^2 \sigma_0 = (x_0^2 +  x_1^2 +  x_2^2 +  x_3^2)\sigma_0.$$ A typical measure on $H$ may take the form
\begin{equation}\label{measure}
d\varsigma(r, \theta,\phi, \psi)= d\tau(r)\, d\theta\, d\Omega(\phi ,\psi )
\end{equation}
with $d\Omega(\phi ,\psi) = \displaystyle{\frac{1}{4\pi}} \,\sin{\phi}\, d\phi \,d\psi .$
Note also that for ${\pu},\qu\in H$, we have $\overline{{\pu}\qu}=\overline{\qu}~\overline{{\pu}}$, $\pu\qu\not=\qu\pu$, $\qu\overline{{ \qu}}=\overline{{\qu}}\qu$, and real numbers commute with quaternions.
%%%%%%%%%%%%%%%%%%%%%%%%%%%%%%%%%%%%%%%%%%%%%%%%%%%%%%%%%%%%%%%%%%%%%%%%%%%%%%%%%%%%%%%%%%%%%%
In defining the position and momentum operators, we shall also need the sliced version of quaternions. We borrow the materials as needed here from \cite{Gen1}.  Let 
\begin{equation}
\mathbb{S}=\{\qu=x_1i+x_2j+x_3k~\vert~x_1,x_2,x_3\in\mathbb{R},~x_1^2+x_2^2+x_3^2=1\},
\end{equation}
we call it a quaternion sphere.
\begin{proposition}\cite{Gen1}\label{Pr1}
For any non-real quaternion $\qu\in H\smallsetminus\mathbb{R}$, there exist, and are unique, $x,y\in\mathbb{R}$ with $y>0$, and $I\in\mathbb{S}$ such that $\qu=x+yI$.
\end{proposition}
\begin{definition}(Slice \cite{Gen1})\label{De1}
For every quaternion $I\in\mathbb{S}$, the complex line $L_I=\mathbb{R}+I\mathbb{R}$ passing through the origin, and containing $1$ and $I$, is called a quaternion slice.
\end{definition}
From the definition we can see that
\begin{equation}\label{Eq1}
H=\bigcup_{I\in\mathbb{S}}L_I\quad\text{and}\quad\bigcap_{I\in\mathbb{S}} L_I=\mathbb{R}.
\end{equation}
One can also easily see that $L_I\subset H$ is commutative, while, elements from two different quaternion slices, $L_I$ and $L_J$ (for $I, J\in\mathbb{S}$ with $I\not=J$), do not necessarily commute.

%%%%%%%%%%%%%%%%%%%%%%%%%%%%%%%%%%%%%%%%%%%%%%%%%%%%%%%%%%%%%%%%%%%%%%%%%%%%%%%%%%%%%%%%%%%%%%%%%%%%%%%%%%%%%%%%%%%%%%%%%%%%%%%%%%%%%%%%%%%%%%%%%%%%%%%%%%%%%%%%%%%%
\subsection{Quaternionic Hilbert spaces}
In this subsection we  define left and right quaternionic Hilbert spaces. For details we refer the reader to \cite{Ad}. We also define the Hilbert space of square integrable functions on quaternions based on \cite{Vis, Gu}.
\subsubsection{Right Quaternionic Hilbert Space}
Let $V_{H}^{R}$ be a linear vector space under right multiplication by quaternionic scalars (again $H$ standing for the field of quaternions).  For $f,g,h\in V_{H}^{R}$ and $\qu\in H$, the inner product
$$\langle\cdot\mid\cdot\rangle:V_{H}^{R}\times V_{H}^{R}\longrightarrow H$$
satisfies the following properties
\begin{enumerate}
\item[(i)]
$\overline{\langle f\mid g\rangle}=\langle g\mid f\rangle$
\item[(ii)]
$\|f\|^{2}=\langle f\mid f\rangle>0$ unless $f=0$, a real norm
\item[(iii)]
$\langle f\mid g+h\rangle=\langle f\mid g\rangle+\langle f\mid h\rangle$
\item[(iv)]
$\langle f\mid g\qu\rangle=\langle f\mid g\rangle\qu$
\item[(v)]
$\langle f\qu\mid g\rangle=\overline{\qu}\langle f\mid g\rangle$
\end{enumerate}
where $\overline{\qu}$ stands for the quaternionic conjugate. We assume that the
space $V_{H}^{R}$ is complete under the norm given above. Then,  together with $\langle\cdot\mid\cdot\rangle$ this defines a right quaternionic Hilbert space, which we shall assume to be separable. Quaternionic Hilbert spaces share most of the standard properties of complex Hilbert spaces. In particular, the Cauchy-Schwartz inequality holds on quaternionic Hilbert spaces as well as the Riesz representation theorem for their duals.  Thus, the Dirac bra-ket notation
can be adapted to quaternionic Hilbert spaces:
$$\mid f\qu\rangle=\mid f\rangle\qu,\hspace{1cm}\langle f\qu\mid=\overline{\qu}\langle f\mid\;, $$
for a right quaternionic Hilbert space, with $\vert f\rangle$ denoting the vector $f$ and $\langle f\vert$ its dual vector. Let $O_R$ be an operator on a right quaternionic Hilbert space. The scalar multiple of $O_R$ should be written as $\qu O_R$ and the action must take the form
\begin{equation}\label{act}
(\qu O_R)\mid f\rangle=(O_R\mid f\rangle)\overline{\qu}.
\end{equation}
The adjoint $O_R^{\dagger}$ of $O_R$ is defined as
\begin{equation}\label{Ad1}
\langle g\mid O_Rf\rangle=\langle O_R^{\dagger}g\mid f\rangle;\quad\text{for all}~~~ f,g\in V_H^R.
\end{equation}
%%%%%%%%%%%%%%%%%%%%%%%%%%%%%%%%%%%%%%%%%%%%%%%%%%%%%%%%%%%%%%%%%%%%%%%%%%%%%%%%%%%%%%%%%%%%%%%%%%%%%%%%%%%%%%%%%%%%%%%%%%%%%%%%%%%%%%%%%%%%%%%%%%%%%%%%%%%%%%%%%%%%%%
\subsubsection{Left Quaternionic Hilbert Space}
Let $V_H^L$ be a linear vector space under left multiplication by quaternionic scalars.  For $f,g,h\in V_{H}^{L}$ and $\qu\in H$, the inner product
$$\langle\cdot\mid\cdot\rangle:V_{H}^{L}\times V_{H}^{L}\longrightarrow H$$
satisfies the following properties
\begin{enumerate}
\item[(i)]
$\overline{\langle f\mid g\rangle}=\langle g\mid f\rangle$
\item[(ii)]
$\|f\|^{2}=\langle f\mid f\rangle>0$ unless $f=0$, a real norm
\item[(iii)]
$\langle f\mid g+h\rangle=\langle f\mid g\rangle+\langle f\mid h\rangle$
\item[(iv)]
$\langle \qu f\mid g\rangle=\qu\langle f\mid g\rangle$
\item[(v)]
$\langle f\mid \qu g\rangle=\langle f\mid g\rangle\overline{\qu}$
\end{enumerate}
Again, we shall assume that the space $V_H^L$ together with $\langle\cdot\mid\cdot\rangle$ is a separable Hilbert space. Also,
\begin{equation}\label{leftcs}
\mid \qu f\rangle=\mid f\rangle\overline{\qu},\hspace{1cm}\langle \qu f\mid=\qu\langle f\mid.
\end{equation}
Note that, because of our convention for inner products, for a left quaternionic Hilbert space, the bra vector $\langle f\mid$ is to be identified with the vector itself, while the ket vector $\mid f \rangle$ is to be identified with its dual. Note also that there is a natural left multiplication by quaternionic scalars on the dual of a right quaternionic Hilbert space and a similar right multiplication on the dual of a left quaternionic Hilbert space.\\
Separable quaternionic Hilbert spaces admit countable orthonormal bases. Let $V_H^R$ be a right quaternionic Hilbert space and let $\{e_v\}_{\nu = 0}^N$  ($N$ could be finite or infinite) be an orthonormal basis for it. Then, $\langle e_\nu \mid e_\mu\rangle = \delta_{\nu \mu}$ and any vector $f \in V_H^R$ has the expansion $f = \sum_\nu e_\nu f_\nu $, with $f_\nu = \langle e_\nu\mid f\rangle \in H$. Using such a basis, it is possible to introduce a multiplication from the left on $V_H^R$ by elements of $H$. Indeed, for $f \in V_H^R$ and $\qu\in H$ we define,
\begin{equation}
  \qu f = \sum_\nu e_\nu(\qu f_\nu ) .
\label{rightmult}
\end{equation}
Further, $\langle \qu f\mid g\rangle=\langle f\mid \overline{\qu}g\rangle$ (see \cite{Ale}).
The field of quaternions $H$ itself can be turned into a left quaternionic Hilbert space by defining the inner product $\langle \qu \mid \qu^\prime \rangle = \qu \qu^{\prime\dag} = \qu\overline{\qu^\prime}$ or into a right quaternionic Hilbert space with  $\langle \qu \mid \qu^\prime \rangle = \qu^\dag \qu^\prime = \overline{\qu}\qu^\prime$. Further note that, due to the non-commutativity of quaternions the sum
\begin{equation}\label{SS1}
\sum_{m=0}^{\infty}\frac{\pu^m\qu^m}{m!}\not=\sum_{m=0}^{\infty}\frac{(\pu\qu)^m}{m!},
\end{equation}
thereby it cannot be written as $\text{exp}(\bf{p}\qu).$ However, in any Hilbert space the norm convergence implies the convergence of the series and
\begin{eqnarray*}\sum_{m=0}^{\infty}\left\vert\frac{\pu^m\qu^m}{m!}\right\vert
&\leq&\sum_{m=0}^{\infty}\frac{|\pu|^m|\qu|^m}{m!}\\
&=&\sum_{m=0}^{\infty}\frac{(|\pu||\qu|)^m}{m!}
=e^{|\bf{p}||\qu|}.
\end{eqnarray*}
Thus the series (\ref{SS1}) converges and, wherever needed, we call it $E(\bf{p},\qu)$.
%%%%%%%%%%%%%%%%%%%%%%%%%%%%%%%%%%%%%%%%%%%%%%%%%%%%%%%%%%%%%%%%%%%%%%%%%%%%%%%%%%%%%%%%%%%%%%%%%%%%%%%%%%%%%%%%%%%%%%%%%%%%%%%%%%%%%%%%%%%%%%%%%%%%%%%%%%%%%%%%%%%%%
\subsubsection{Quaternionic Hilbert Spaces of Square Integrable Functions}
Let $(X, \mu)$ be a measure space and $H$  the field of quaternions, then
$$\left\{f:X\rightarrow H \left|  \int_X|f(x)|^2d\mu(x)<\infty \right.\right\}$$\label{L^2}\noindent
is a right quaternionic Hilbert space which is denoted by $L^2_H(X,\mu)$, with the (right) scalar product
\begin{equation}
\langle f \mid g\rangle =\int_X\overline{ f(x)}{g(x)} d\mu(x),
\label{left-sc-prod}
\end{equation}
where $\overline{f(x)}$ is the quaternionic conjugate of $f(x)$, and (right)  scalar multiplication $fa, \; a\in H,$ with $(fa)(q) = f(q)a$ (see \cite{Gu,Vis} for details). Similarly, one could define a left quaternionic Hilbert space of square integrable functions.
%%%%%%%%%%%%%%%%%%%%%%%%%%%%%%%%%%%%%%%%%%%%%%%%%%%%%%%%%%%%%%%%%%%%%%%%%%%%%%%%%%%%%%%%%%%%%%%%%%%%%%%%%%%%%%%%%%%%%%%%%%%%%%%%%%%%%%%%%%%%%%%%%%%%%%%%%%%%%%%%%%%%%%
\section{Coherent states on right quaternion Hilbert spaces}\label{CSLQH}
The main content of this section is extracted from \cite{Thi2} as needed here. For an enhanced explanation we refer the reader to \cite{Thi2}. In \cite{Thi2} the authors have defined coherent states on $V_{H}^{R}$ and $V_{H}^{L}$, and also established the normalization and resolution of the identities for each of them. We briefly revisit the coherent states of $V_{H}^{R}$ and the normalization and resolution of the identity.
Let $\{\mid f_{m}\rangle\}_{m=0}^{\infty}$ be an orthonormal basis of $V_{H}^{R}$. For $\qu\in V_{H}^{R}$, the coherent states are defined as vectors in $V_{H}^{R}$ in the form of
\begin{equation}\label{CS}
\mid\qu\rangle=\mathcal N (\mid\qu\mid)^{-\frac{1}{2}}\sum_{m=0}^{\infty}\mid f_m\rangle \frac{\qu^{m}}{\sqrt{\rho(m)}},
\end{equation}
where $\mathcal N (\mid\qu\mid)$ is the normalization factor and $\{\rho(m)\}_{m=0}^{\infty}$ is a positive sequence of real numbers. Using conditions (\ref{san1}), we can determine the normalization factor $\mathcal N (\mid\qu\mid)$, and the resolution of the identity. In order for the norm of $\mid\qu\rangle$ to be finite, we must have
\begin{equation}
\langle\qu\mid\qu\rangle=\mathcal N (\mid\qu\mid)^{-1}\sum_{m=0}^{\infty}\frac{r^{2m}}{\rho(m)}<\infty.
\end{equation}
Therefore, if the positive sequence $\{\rho(m)\}_{m=0}^{\infty}$ of real numbers converges to $\ell>0$, then we are required to restrict the domain to  
\begin{equation}\label{Dom}
\mathcal D=[0,\sqrt{\ell}\,)\times[0,\pi]\times[0,2\pi)^{2}
\end{equation}
so that the convergence of the above series is guaranteed. The typical measure (\ref{measure}) is an appropriate one on the domain $\mathcal D$ too. By requiring $\langle\qu\mid\qu\rangle=1$, the normalization factor is obtained as
\begin{equation}
\mathcal N (\mid\qu\mid)=\sum_{m=0}^{\infty}\frac{r^{2m}}{\rho(m)}.
\end{equation}
Using the measure $d\varsigma(r, \theta,\phi, \psi)$ one can obtain the following operator valued integral on the domain $\mathcal D$ of (\ref{Dom}):
\begin{equation}
\int_{\mathcal D}\mid\qu\rangle\langle\qu\mid d\varsigma(r, \theta,\phi, \psi)=\sum_{m=0}^{\infty}\frac{2\pi}{\rho(m)}\displaystyle \int_{0}^{\sqrt{\ell}}\frac{r^{2m}}{\mathcal N (\mid\qu\mid)}\mid f_{m}\rangle\langle f_{m}\mid d\tau(r),
\end{equation}
and in obtaining it we have used the identity
\begin{equation}\label{int}
\int_{0}^{2\pi}\int_0^{\pi}\int_0^{2\pi} e^{i(m-l)\theta\sigma(\hat{n})}\sin\phi\, d\phi\, d\theta\, d\varphi=2\pi\delta_{ml}\mathbb{I}_{2},
\end{equation}
%\noindent
where $\delta_{ml}$ is the Kronecker delta. The resolution of the identity, 
\begin{equation}\label{res}
\int_{\mathcal D}\mid\qu\rangle\langle\qu\mid d\varsigma(r, \theta,\phi, \psi)=\mathbb{I}_{V_{H}^{R}},
\end{equation}
where $\mathbb{I}_{V_{H}^{R}}$ is the identity operator on $V_{H}^{R}$, is obtained if there is a measure to satisfy the moment problem, 
\begin{equation}\label{EE1}
\frac{2\pi}{\rho(m)}\int_{0}^{\sqrt{\ell}}\frac{r^{2m}}{\mathcal N (\mid\qu\mid)}d\tau(r)\mathbb{I}_{2}=\mathbb{I}_{2}.
\end{equation}
 If the measure $d\tau(r)$ is chosen such that
\begin{equation}\label{eaux}
d\tau(r)=\frac{\mathcal N (\mid\qu\mid)}{2\pi}\lambda(r)dr,
\end{equation}
then there exists an auxiliary density $\lambda(r)$ to solve (\ref{EE1}), that is, we get
\begin{equation}
\int_{0}^{\sqrt{\ell}}r^{2m}\lambda(r)dr\mathbb{I}_{2}=\rho(m)\mathbb{I}_{2}.
\end{equation}
Particularly, if $\rho(m)=m!$, then the normalization factor $\mathcal N (\mid\qu\mid)=e^{|\qu|^{2}}$ and $\ell=\infty$. The resolution of the identity can be established for (\ref{CS}) with $\lambda(r)=2re^{-r^{2}}$. In this case $\mathcal{D}=H$ and the CS are called {\em right quaternionic canonical coherent states}. For the purpose of quantizing the quaternions we shall use this canonical set of CS.
%%%%%%%%%%%%%%%%%%%%%%%%%%%%%%%%%%%%%%%%%%%%%%%%%%%%%%%%%%%%%%%%%%%%%%%%%%%%%%%%%%%%%%%%%%%%%%%%%%%%%%%%%%%%%%%%%%%%%%%%%%%%%%%%%%%%%%%%%%%%%%%%%%%%%%%%%%%%%%%%%%%%%%%
\section{Coherent state quantization: General scheme}
Let $(X,\mu)$ be a measure space and $L^2(X,\mu)$ be given by
$$\left\{f:X\rightarrow \mathbb{C}~\vert~\int_X|f(x)|^2d\mu(x)<\infty\right\}.$$
The Berezin-Toeplitz or anti-Wick or coherent state quantization, as used by various authors in the literature, associates a classical observable that is a function $f(x)$ on $X$ to an operator valued integral. We continue with the general procedure described in \cite{Gaz} and applied, for example, in \cite{Nic,Ga, AGH}.\\
Choose a countable orthonormal family $$\mathcal{O}=\{\phi_n~\vert~n=0,1,2\cdots\}$$ in $L^2(X,\mu)$, that is
\begin{equation}\label{E1}
\langle\phi_n|\phi_m\rangle=\int_X\overline{\phi_n(x)}{\phi_m(x)}d\mu(x)=\delta_{mn},
\end{equation}
and assume that
\begin{equation}\label{E2}
0<\sum_{n=0}^{\infty}|\phi_n(x)|^2:=\mathcal{N}(x)<\infty\quad a.e.
\end{equation}
holds. Let $\mathfrak{H}$ be a separable complex Hilbert space with orthonormal basis $\{|e_n\rangle~\vert~n=0,1,2\cdots\}$ in 1-1 correspondence with $\mathcal{O}$. In particular $\mathfrak{H}$ can be taken as $\mathfrak{H}=\overline{\text{span}\mathcal{O}}$ in $L^2(X,\mu)$, where the bar stands for the closure. Then the family $\mathcal{F}_{\mathfrak{H}}=\{|x\rangle~\vert~x\in X\}$ with
\begin{equation}\label{E3}
|x\rangle=\mathcal{N}(x)^{-\frac{1}{2}}\sum_{n=0}^{\infty}\overline{\phi_n(x)}|e_n\rangle\in\mathfrak{H}
\end{equation}
forms a set of coherent states(CS). From (\ref{E1}) and (\ref{E2}) we have
\begin{eqnarray}
& &\langle x\vert x\rangle=1\label{E4}\\
& &\int_X\mathcal{N}(x)|x\rangle\langle x|d\mu(x)=\mathbb{I}_{\mathfrak{H}},\label{E5}
\end{eqnarray}
where $\mathbb{I}_{\mathfrak{H}}$ is the identity operator on $\mathfrak{H}$. We call the set $\mathcal{F}_{\mathfrak{H}}$ a set of CS only for satisfying the normalization and a resolution of the identity. Equation (\ref{E5}) allows us to implement CS or frame quantization of the set of parameters $X$ by associating a function
$$X\ni x\mapsto f(x)$$
that satisfies appropriate conditions with the following operator in $\mathfrak{H}$
\begin{equation}\label{E6}
f(x)\mapsto A_f=\int_X\mathcal{N}(x)f(x)|x\rangle\langle x|d\mu(x).
\end{equation}
The matrix elements of $A_f$ with respect to the basis $\{|e_n\rangle\}$ are given by
\begin{eqnarray*}\label{E7}
\left(A_f\right)_{mn}&=&\langle e_m\vert A_f\vert e_n\rangle\\%\nonumber
&=&\int_Xf(x)\overline{\phi_m(x)}\phi_n(x)d\mu(x).
\end{eqnarray*}
The operator $A_f$ is
\begin{enumerate}
\item[(a)]symmetric if $f(x)$ is real valued.
\item[(b)]bounded if $f(x)$ is bounded.
\item[(c)]self-adjoint if $f(x)$ is real semi-bounded (through Friedrich's extension).
\end{enumerate}
In order to view the upper symbol $f$ of $A_f$ as a quantizable object (with respect to the family $\mathcal{F}_{\mathfrak{H}}$) a reasonable requirement is that the so-called lower symbol of $A_f$ defined as
\begin{eqnarray*}\label{E8}
\check{f}(x)&=&\langle x\vert A_f\vert x\rangle\\&=&\int_X\mathcal{N}(x')f(x')|\langle x\vert x'\rangle|^2d\mu(x')
\end{eqnarray*}
be a smooth function on $X$ with respect to some topology assigned to the set $X$. Associating to the classical observable $f(x)$ the mean value $\langle x|A_f|x\rangle$ one can also get the so-called Berezin transform $B[f]$ with
$B[f](x)=\langle x|A_f|x\rangle$, for example, see \cite{Mo} for details.
%%%%%%%%%%%%%%%%%%%%%%%%%%%%%%%%%%%%%%%%%%%%%%%%%%%%%%%%%%%%%%%%%%%%%%%%%%%%%%%%%%%%%%%%%%%%%%%%%%%%%%%%%%%%%%%%%%%%%%%%%%%%%%%%%%%%%%%%%%%%%%%%%%%%%%%%%%%%%%%%%%%%%
\section{Quantization of the quaternions}
In this section we shall adapt the general procedure outlined in the above section to quaternions.
Since $(H,d\varsigma(r, \theta,\phi, \psi))$ is a measure space, the set 
\begin{equation*}
\left\{f:H\rightarrow H\mid\int_H|f(\qu)|^2d\varsigma(r, \theta,\phi, \psi)<\infty\right\}
\end{equation*}
is a space of right quaternionic square integrable functions and is denoted by  $L^2_{H}(H,d\varsigma(r, \theta,\phi, \psi))$. Define the sequence of functions $\{\phi_{n}\}_{n=0}^{\infty}$ such that $$\phi_{n}: H\longrightarrow H$$ by 
\begin{equation}\label{phi}
\phi_{n}(\qu)=\frac{\overline{\qu}^{n}}{\sqrt{n!}},~~\mbox{~~for all~~}~~\qu\in H.\end{equation}
Then $\phi_{n}\in L^2_{H}(H,d\varsigma(r, \theta,\phi, \psi))$, for all $n=0,1,2\cdots$ and from (\ref{int}) $\langle\phi_{m}\mid\phi_{n}\rangle=\delta_{mn}$ (see \cite{Thi2}).
That is, $$\mathcal{O}=\{\phi_n~\vert~n=0,1,2\cdots\}$$ is an orthonormal set in $L^2_{H}(H,d\varsigma(r, \theta,\phi, \psi))$. The right quaternionic span of $\mathcal{O}$ is the space of anti-right-regular functions \cite{Thi1} (the counterpart of complex anti-holomorphic functions). Let $\mathfrak{H}$ be a separable right quaternionic Hilbert space with an orthonormal basis  $$\mathcal{E}=\{~\mid e_{n}\rangle~\mid~ n=0,1,2\cdots~\}$$ which is in $1-1$ correspondence with $\mathcal{O}$. Then the coherent states (\ref{CS}) become
\begin{equation}\label{CS1}
\mid\gamma_{\qu}\rangle=e^{-\mid\qu\mid^{2}/2}\sum_{m=0}^{\infty}\mid e_m\rangle\overline{\phi_{m}}(\qu).
\end{equation}
Using the set of CS (\ref{CS1}) we shall establish the coherent state quantization on $\mathfrak{H}$ by associating a function $$H\ni \qu\longmapsto f(\qu,\overline{\qu}).$$
Now let us define the operator on $\mathfrak{H}$ by
\begin{equation}
f(\qu,\overline{\qu})\mapsto A_{f},
\end{equation}
where $A_{f}$ is given by the operator valued integral
\begin{equation}\label{Qmap}
A_{f}=\int_{H}\mid\gamma_{\qu}\rangle f(\qu,\overline{\qu})\langle\gamma_{\qu}\mid  d\varsigma(r, \theta,\phi, \psi).
\end{equation}

Now 
\begin{eqnarray*}
A_{f}&=&\int_{H}\mid\gamma_{\qu}\rangle f(\qu,\overline{\qu})\langle\gamma_{\qu}\mid d\varsigma(r, \theta,\phi, \psi)\\
&=&\sum_{m=0}^{\infty}\sum_{l=0}^{\infty}\frac{\mid e_{m}\rangle J_{m,l}\langle e_{l}\mid}{\sqrt{m!~l!}};
\end{eqnarray*}
where the integral $J_{m,l} $ is given by $$\displaystyle \iiiint\limits_{[0,\infty)\times[0,\pi]\times[0,2\pi)^{2}}\frac{ {\qu^{m}}f(\qu,\overline{\qu}){\overline{\qu}^{l}}}{e^{r^2}}d\varsigma(r, \theta,\phi, \psi).$$
By direct calculation we have that 
if  $f(\qu,\overline{\qu})=\qu$, then
\begin{equation}\label{Aq}
A_{\qu}
=\sum_{m=0}^{\infty}\sqrt{(m+1)}\mid e_{m}\rangle\langle e_{m+1}\mid
\end{equation}
and if $f(\qu,\overline{\qu})=\overline{\qu}$, then
\begin{equation}\label{Aqb}
A_{\overline{\qu}}
=\sum_{m=0}^{\infty}\sqrt{(m+1)}\mid e_{m+1}\rangle\langle e_{m}\mid.
\end{equation}
Moreover if $f(\qu,\overline{\qu})=1$, then 
$A_{1}
=\mathbb{I}_{\mathfrak{H}}.$
It should be mentioned that, since the operator $A_{f}$ is a quaternionic operator, the usual properties of its complex counterpart may not hold. In this regard, each property used must be validated. First let $|f\rangle, |g\rangle\in\mathfrak{H}$. Since $\mathfrak{H}$ is a right Hilbert space, there are scalars $\{\alpha_l\}, \{\beta_j\}$ in $H$ such that
$$|f\rangle=\sum_{l=0}^{\infty}|e_l\rangle\alpha_l\quad\text{and}\quad
|g\rangle=\sum_{j=0}^{\infty}|e_j\rangle\beta_j.$$
With these it can be seen that
\begin{eqnarray*}
\langle A_{\overline{\qu}}g\mid f\rangle&=&\langle g\mid A_{\qu}f\rangle
=\sum_{m=0}^{\infty}\overline{\beta}_m\alpha_{m+1}\sqrt{m+1}.
\end{eqnarray*}
That is,  
$$\langle A_{\overline{\qu}}g\mid f\rangle=\langle g\mid A_{\qu}f\rangle;\quad\text{for all}\quad |f\rangle, |g\rangle\in\mathfrak{H}.$$
Hence $A_{\overline{\qu}}$ is the adjoint of $A_{\qu}$ and vice-versa. Now $A_{f}$ is an operator from $\mathfrak{H}$ to $\mathfrak{H}$, and if $\mathfrak{H}=\overline{span\mathcal{O}}$ (right linear span over $H$), then it is a subspace of $L^2_{H}(H,d\varsigma(r, \theta,\phi, \psi))$. That is,
$$A_{f}:\mathfrak{H}\longrightarrow \mathfrak{H}~~\mbox{~~by}\quad 
A_{f}(u)=A_{f}\mid u\rangle,$$ for all $u\in \mathfrak{H}$. Hence, $A_{f}(u)$ will be determined by the integral
$$\int_{H}\mid\gamma_{\qu}\rangle f(\qu,\overline{\qu})\langle\gamma_{\qu}\mid u\rangle d\varsigma(r, \theta,\phi, \psi).$$
Moreover, for each $u\in\mathfrak{H},~~A_{f}\mid u\rangle\in\mathfrak{H}$. For $|u\rangle, |v\rangle\in\mathfrak{H}$, it can also be considered as a function
$$A_{f}:\mathfrak{H}\times \mathfrak{H}\longrightarrow H~~\mbox{~~by}\quad 
A_{f}(u,v)=\langle u\mid A_{f}\mid v\rangle.$$
Thereby, $A_{f}(u,v)$ will be determined by the quaternion valued integral 
$$\int_{H}\langle u\mid\gamma_{\qu}\rangle f(\qu,\overline{\qu})\langle\gamma_{\qu}\mid v\rangle d\varsigma(r, \theta,\phi, \psi).$$
Since $\mid\gamma_{\qu}\rangle$ is a column vector and $\langle\gamma_{\qu}\mid$ is a row vector, we can see that the operator $A_{f}$ is a matrix and the matrix elements with respect to the basis $\{\mid e_{n}\rangle\}$ are given by
$$(A_{f})_{mn}=\langle e_{m}\mid A_{f}\mid e_{n}\rangle.$$ That is, $(A_{f})_{mn}$ is determined by the integral $$\int_{H}\langle e_{m}\mid\gamma_{\qu}\rangle f(\qu,\overline{\qu})\langle\gamma_{\qu}\mid e_{n}\rangle d\varsigma(r, \theta,\phi, \psi).$$
We have
$$\langle e_{m}\mid\gamma_{\qu}\rangle=\mathcal N (\mid\qu\mid)^{-\frac{1}{2}}~\overline{\phi_{m}(\qu)}$$ and$$\langle\gamma_{\qu}\mid e_{n}\rangle=\overline{\langle e_{n}\mid\gamma_{\qu}\rangle}=\mathcal N (\mid\qu\mid)^{-\frac{1}{2}}~{\phi_{n}(\qu)}.$$
Therefore
$$(A_{f})_{mn}=\int_{H}\mathcal N (\mid\qu\mid)^{-1}\overline{\phi_{m}(\qu)}f(\qu,\overline{\qu})\phi_{n}(\qu). d\varsigma(r, \theta,\phi, \psi).$$ 
Hence, it can easily be seen that
\begin{eqnarray*}
(A_{\qu})_{k,l}&=&\langle e_k|A_{{\qu}}|e_l\rangle=\left\{\begin{array}{ccc}
\sqrt{k+1}&\text{if}&l=k+1\\
0&\text{if}&l\not=k+1,\end{array}\right.\\
(A_{\overline{\qu}})_{k,l}&=&\langle e_k|A_{\overline{\qu}}|e_l\rangle=\left\{\begin{array}{ccc}
\sqrt{k}&\text{if}&l=k-1\\
0&\text{if}&l\not=k-1.\end{array}\right.
\end{eqnarray*}

%%%%%%%%%%%%%%%%%%%%%%%%%%%%%%%%%%%%%%%%%%%%%%%%%%%%%%%%%%%%%%%%%%%%%%%%%%%%%%%%%%%%%%%%%%%%%%%%%%%%%%%%%%%%%%%%%%%%%%%%%%%%%%%%%%%%%%%%%%%%%%%%%%%%%%%%%%%%%%%%%%%%%%%%%%%%%%%%%%%%%%%%%%%%%%%%%%%%%%%%
\noindent
Let us realize the operator $A_f$ as annihilation and creation operators. From (\ref{Aq}) and (\ref{Aqb}) we have $A_{{\qu}}\mid e_{0}\rangle=0\,$,
$$A_{{\qu}}\mid e_{m}\rangle=\sqrt{m}\mid e_{m-1}\rangle\,; ~ m=1,2,\cdots$$ and 
$$A_{\overline{\qu}}\mid e_{m}\rangle=\sqrt{m+1}\mid e_{m+1}\rangle\,; ~ m=0,1,2,\cdots$$
That is, $A_{{\qu}},A_{\overline{\qu}}$ are annihilation and creation operators respectively. Moreover, one can easily see that $A_{\qu}\mid\gamma_{\qu}\rangle=\mid\gamma_{\qu}\rangle \qu$, which is in complete analogy with the action of the annihilation operator on the ordinary harmonic oscillator CS and the result obtained in \cite{Thi2}. We can also write
$$\mid e_n\rangle=\frac{(A_{\overline{\qu}})^n}{\sqrt{n!}}\mid e_0\rangle.$$
Further, real numbers commute with quaternions. Therefore according to (\ref{act}), for example, we have
$$(\overline{\qu}A_{\overline{\qu}})^2\mid e_0\rangle=(\overline{\qu}A_{\overline{\qu}})(\overline{\qu}A_{\overline{\qu}})|e_0\rangle
=(\overline{\qu}A_{\overline{\qu}})|e_1\rangle\sqrt{1}\qu
=|e_2\rangle \sqrt{2}\sqrt{1}~\qu^2=|e_2\rangle\sqrt{2!}~\qu^2.$$
Accordingly,
$$\mid\gamma_{\qu}\rangle=e^{-|\qu|^2/2}e^{\overline{\qu}A_{\overline{\qu}}}|e_0\rangle.$$
For quaternionic exponentials we refer the reader to \cite{Eb} (pp 204).
Now a direct calculation shows that
\begin{eqnarray*}
A_{{\qu}}A_{\overline{\qu}}
&=&\sum_{m=0}^{\infty}(m+1)\mid e_{m}\rangle\langle e_{m}\mid
\end{eqnarray*}
and
\begin{eqnarray*}
A_{\overline{\qu}}A_{{\qu}}
&=&\sum_{m=0}^{\infty}(m+1)\mid e_{m+1}\rangle\langle e_{m+1}\mid.
\end{eqnarray*}
Therefore the commutator of $A_{\overline{\qu}},A_{{\qu}}$ takes the form
\begin{eqnarray*}
[A_{{\qu}},A_{\overline{\qu}}]&=&A_{{\qu}}A_{\overline{\qu}}-A_{\overline{\qu}}A_{{\qu}}\\
&=&\sum_{m=0}^{\infty}\mid e_{m}\rangle\langle e_{m}\mid
=\mathbb{I}_{\mathfrak{H}}.
\end{eqnarray*}
\begin{remark}
The operator $A_f$ in (\ref{Qmap}) is formed by the vector $\mid\gamma_{\qu}\rangle f(\qu,\overline{\qu})$, which is the right scalar multiple of the vector $\mid\gamma_{\qu}\rangle$ by the scalar $ f(\qu,\overline{\qu})$, and the dual vector $\langle\gamma_{\qu}\mid$. Instead if one takes
\begin{equation}\label{Af}
A_f=\int_H f(\qu,\overline{\qu})\mid\gamma_{\qu}\rangle\langle\gamma_{\qu}\mid d\varsigma(r, \theta,\phi, \psi),
\end{equation}
then it is formed by $ f(\qu,\overline{\qu})\mid\gamma_{\qu}\rangle$ (a left scalar multiple of a right Hilbert space vector) and the dual vector $ \langle\gamma_{\qu}\mid$, which is  unconventional . Further, due to the non-commutativity of quaternions, an $A_f$ in the form (\ref{Af}) would have caused severe technical problems in the follow up computations.
\end{remark}
%%%%%%%%%%%%%%%%%%%%%%%%%%%%%%%%%%%%%%%%%%%%%%%%%%%%%%%%%%%%%%%%%%%%
\subsection{Number, position and momentum operators and Hamiltonian}
Let $N=A_{\overline{\qu}}A_{\qu}$, then we have
\begin{eqnarray*}
N\mid e_k\rangle&=&A_{\overline{\qu}}A_{\qu}\mid e_k\rangle\\
&=&\sum_{m=0}^{\infty}\mid e_{m+1}\rangle\langle e_{m+1}\mid e_k\rangle(m+1)\\
&=&\mid e_k\rangle k.
\end{eqnarray*}
Thereby $N$ acts as the number operator and the Hilbert space $\mathfrak{H}$ is the quaternionic Fock space (for quaternion Fock spaces see \cite{Al}). As an analogue of the usual harmonic oscillator Hamiltonian, if we take $\mathcal{H}_h=N+\mathbb{I}_{\mathfrak{H}}$, then  $\mathcal{H}_h\mid e_n\rangle=\mid e_n\rangle (n+1)$, which is a Hamiltonian in the right quaternionic Hilbert space $\mathfrak{H}$ with spectrum $(n+1)$ and eigenvector $\mid e_n\rangle$.

Following the complex formalism (see remark (\ref{Rem2})), for $\qu\in H$ if we take $\displaystyle\mathfrak{q}=\frac{1}{\sqrt{2}}(\qu+\overline{\qu})$, then we can have a self-adjoint position operator as $$\displaystyle Q=\frac{1}{\sqrt{2}}(A_{\qu}+A_{\overline{\qu}}).$$
Now let us turn our attention to the momentum operator. In the case of the momentum operator, the complex formalism does not transfer to quaternions.
In the case of quaternions we have three imaginary units, $i, j$ and $k$, and if we try to duplicate the complex momentum coordinate with one of $i, j$ or $k$, that is, if we take
$$\mathfrak{p}=\frac{-i}{\sqrt{2}}(\qu-\overline{\qu}),$$
then the operator $P$ becomes
$$P=\frac{-i}{\sqrt{2}}(A_{\qu}-A_{\overline{\qu}})$$
and due to the non-commutativity of quaternions $P$ is not self-adjoint (see remark \ref{rema1}). Further, a simple calculation shows that, the analogue of the complex operator $H_c$ in remark (\ref{Rem2}) is $H_h=\frac{1}{2}(\mathfrak{q}^2+\mathfrak{p}^2)\not=|\qu|^2$. However, the lower symbol of $N$ is $\langle \gamma_{\qu}\mid N\mid\gamma_{\qu}\rangle=|\qu|^2$ and through a rather lengthy calculation we can see that  $A_{|\qu|^2}=N+\mathbb{I}_{\mathfrak{H}}$. 

In fact, according to \cite{Ad}, there is no quaternion self-adjoint momentum operator that has all the properties of the momentum expected by analogy with the momentum operator in complex quantum mechanics. For various attempts and their drawbacks one can see \cite{Ad} (pages 52-64).
However, if we restrict ourselves to a quaternion slice, then we can have self-adjoint position and momentum operators with all the expected properties of their complex counterparts.
In order to exhibit this, let us see the structure of CS on a slice.
\begin{enumerate}
\item[$\bullet$] Since elements in a quaternion slice commute, a quaternion slice is isomorphic to the complex plane. That is, for each $I\in\mathbb{S}$, $L_I$ is isomorphic to $\mathbb{C}$.
\item[$\bullet$]While we are on a slice, $L_I$, the set of CS is formed with elements from the slice $L_I$ and the CS belongs to the right quaternionic Hilbert space over the field $L_I$  and we denote this Hilbert space by $\mathfrak{H}_{L_I}$. 
\item[$\bullet$] Let $\qu_I\in L_I$, $\qu_I=re^{I\theta}; r>0, 0\leq\theta<2\pi$, then the normalization factor of the CS, over the slice $L_I$, is given by $\mathcal{N}(\qu_I)=e^{|\qu_I|^2}$ and a resolution of the identity is obtained with the measure $d\mu_I(r,\theta)=\frac{1}{2\pi}re^{-r^2}drd\theta$.
\end{enumerate}
Even though a quaternion slice is isomorphic to $\C$, for the sake of completeness, while we are on a slice, let us demonstrate few facts. For $\qu_I\in L_I$,
let us define the position and momentum coordinates by
$$\mathfrak{q}_I=\frac{1}{\sqrt{2}}(\qu_I+\overline{\qu}_I)~~\mbox{~~and~~}~~
\mathfrak{p}_I=\frac{-I}{\sqrt{2}}(\qu_I-\overline{\qu}_I),$$
then, since commutativity holds among $I, \qu$ and $\overline{\qu}$, the Hamiltonian can be calculated as $$\displaystyle H_I=\frac{1}{2}\left(\mathfrak{q}_I^{2}+\mathfrak{p}_I^{2}\right)=|\qu_I|^2.$$
Recall that on a right quaternionic Hilbert space operators are multiplied on the left by quaternion scalars. From the position and momentum coordinates, using linearity, we get the position operator, $Q_I$, and the momentum operator, $P_I$, as
$$Q_I=\frac{1}{\sqrt{2}}\left(A_{\qu_I}+A_{\overline{\qu}_I}\right)~~\mbox{~~and~~}~~
P_I=\frac{-I}{\sqrt{2}}\left(A_{\qu_I}-A_{\overline{\qu}_I}\right).$$
Since $(A_{\overline{\qu}_I})^{\dagger}=A_{\qu_I}$ and $(-I)^{\dagger}=I$, the operators $P_I$ and $Q_I$ are self-adjoint. Using the fact (\ref{act}) we can see that $A_{\overline{\qu}_I}(I A_{\qu_I})=I A_{\overline{\qu}_I}A_{\qu_I}$. With the aid of this we get
\begin{eqnarray*}
Q_IP_I&=&
=-\frac{1}{2}I\,[{A_{\qu_I}}^{2}+A_{\overline{\qu}_I}A_{\qu_I}-A_{\qu}A_{\overline{\qu}_I}-{A_{\overline{\qu}_I}}^{2}]
 \end{eqnarray*}
and
\begin{eqnarray*}
P_IQ_I&=&
=-\frac{1}{2}I\,[{A_{\qu_I}}^{2}-A_{\overline{\qu}_I}A_{\qu_I}+A_{\qu_I}A_{\overline{\qu}_I}-{A_{\overline{\qu}_I}}^{2}].
 \end{eqnarray*}
Thereby we have the commutator $$[Q_I,P_I]=Q_IP_I-P_IQ_I=I\,[A_{\qu_I},A_{\overline{\qu}_I}]=I\mathbb{I}_{\mathfrak{H}_{L_I}}.$$
We also have
\begin{eqnarray*}
Q_I^{2}&=&
\frac{1}{2}\,[{A_{\qu_I}}^{2}+A_{\overline{\qu}_I}A_{\qu_I}+A_{\qu_I}A_{\overline{\qu}_I}
+{A_{\overline{\qu}_I}}^{2}]\quad\text{and}\\
P_I^{2}&=&
-\frac{1}{2}\,[{A_{\qu_I}}^{2}-A_{\overline{\qu}_I}A_{\qu_I}-A_{\qu_I}A_{\overline{\qu}_I}
+{A_{\overline{\qu}_I}}^{2}]
 \end{eqnarray*}
Hence $$\hat{H}_I=\frac{Q_I^{2}+P_I^{2}}{2}=\frac{1}{2}[A_{\overline{\qu}_I}A_{\qu_I}+A_{\qu_I}A_{\overline{\qu}_I}]
=A_{\overline{\qu}_I}A_{\qu_I}+\frac{1}{2}[A_{\qu_I}A_{\overline{\qu}_I}-A_{\overline{\qu}_I}A_{\qu_I}]
=N_I+\frac{1}{2}\mathbb{I}_{\mathfrak{H}_{L_I}},$$
which is in complete analogy with the complex case in the sense of {\em canonical quantization}, which simply replaces the classical coordinates by quantum observables (corresponding self-adjoint operators). 
\begin{remark}\label{Rem2}
In the complex quantum mechanics, for the canonical CS $\mid z\rangle,~~z\in \mathbb{C}$, the lower symbol or the expectation value of the number operator, $\langle z|N|z\rangle$, is precisely $|z|^2$. The position and momentum coordinates are $q=\frac{1}{\sqrt{2}}(z+\overline{z})$ and $p=\frac{-i}{\sqrt{2}}(z-\overline{z})$ and by linearity one infers that the position and momentum operators as $Q=\frac{1}{\sqrt{2}}(A_z+A_{\overline{z}})$ and $P=\frac{-i}{\sqrt{2}}(A_z-A_{\overline{z}})$. The CS quantized classical harmonic oscillator, $H_c=\frac{1}{2}(q^2+p^2)$, is $A_{H_c}=A_{|z|^2}=N+\mathbb{I}_{\mathfrak{H}_c}$, where $\mathbb{I}_{\mathfrak{H}_c}$ is the identity operator of the complex Fock space $\mathfrak{H}_c$. The operators $Q$ and $P$ satisfy the commutation rule $[Q,P]=i\mathbb{I}_{\mathfrak{H}_c}$ and are self-adjoint. If one simply takes the canonical quantization of the classical Hamiltonian it becomes $\hat{H}_c=\frac{1}{2}(Q^2+P^2)=N+\frac{1}{2}\mathbb{I}_{\mathfrak{H}_c}$. For details we refer the reader to \cite{Nic, Gaz}.
\end{remark}
\begin{remark}\label{rema1}
For a complex scalar $\alpha\in\mathbb{C}$ and an operator $T$ in a complex Hilbert space, the adjoint of the scalar multiple, $\alpha T$, is taken as $(\alpha T)^{\dagger}=\overline{\alpha}T^{\dagger}$.  However, in general, this is not true for a non-real quaternionic scalar multiple of an operator on a quaternionic Hilbert space. The following simple example validates this claim.
\begin{example}\label{Exa} The quaternion field $H$ with the inner product $\langle \pu|\qu\rangle=\overline{\pu}\qu;~~\pu,\qu\in H,$ is a right quaternionic Hilbert space. Then by (\ref{act}) the identity operator, $\mathfrak{I}$, on $H$ is self-adjoint. For a fixed $\alpha\in H\setminus\mathbb{R}$, if $(\alpha\mathfrak{I})^{\dagger}=\overline{\alpha}~\mathfrak{I}^{\dagger}$, then, keeping (\ref{Ad1}) in mind, for $\pu,\qu\in H$, we have
$$\langle\pu|(\alpha\mathfrak{I})(\qu)\rangle
=\langle\pu|\mathfrak{I}(\qu)\overline{\alpha}\rangle
=\langle\pu|\qu\overline{\alpha}\rangle
=\overline{\pu}\qu\overline{\alpha}$$
and
$$\langle (\alpha\mathfrak{I})^{\dagger}(\pu)|\qu\rangle
=\langle (\overline{\alpha}\mathfrak{I}^{\dagger})(\pu)|\qu\rangle
=\langle (\overline{\alpha}\mathfrak{I})(\pu)|\qu\rangle
=\langle \mathfrak{I}(\pu)\alpha|\qu\rangle
=\langle \pu\alpha|\qu\rangle
=\overline{\pu\alpha}\qu=\overline{\alpha}~\overline{\pu}\qu.$$
In particular, for example, if $\alpha=i+2j,~\qu=j,~\pu=k,$ then we get
$$\langle\pu|(\alpha\mathfrak{I})(\qu)\rangle=1-2k\quad\text{and}\quad
\langle (\alpha\mathfrak{I})^{\dagger}(\pu)|\qu\rangle=1+2k.$$
Therefore $(\alpha\mathfrak{I})^{\dagger}\not=\overline{\alpha}\mathfrak{I}^{\dagger}$.
\end{example}
In general, following the steps of the above example we can easily see that if $T$ is an operator on a right quaternionic Hilbert space $V_H^R$ and $\alpha\in H\setminus\mathbb{R} $, then to hold $(\alpha T)^{\dagger}=\overline{\alpha}T^{\dagger}$ we need
$$\overline{\alpha}\langle g|T(f)\rangle=\langle g|T(f)\rangle\overline{\alpha};\quad f,g\in V_H^R,$$ which is not necessarily true in quaternions. That is, the commutativity of the field on which the Hilbert space is formed guarantees the property $(\alpha T)^{\dagger}=\overline{\alpha}T^{\dagger}$ as $\alpha$ belongs to the same field.
 Since real numbers commute with quaternions, if $T$ is an operator on a quaternionic Hilbert space, then for $\alpha\in\mathbb{R}$, we have $(\alpha T)^{\dagger}=\alpha T^{\dagger}$, see \cite{Gi}. 
\end{remark}
%%%%%%%%%%%%%%%%%%%%%%%%%%%%%%%%%%%%%%%%%%%%%%%%%%%%%%%%%%%%%%%%%%%%%%%%%%%%%
\subsection{Oscillator algebra}
From now on we are back to the general situation, that is, we are not on a slice anymore. Therefore, as before a simple calculation shows that
\begin{eqnarray*}
NA_{\qu}&=&\sum_{m=0}^{\infty}m\sqrt{m+1}\mid e_m\rangle\langle e_{m+1}\mid,\\
A_{\qu}N&=&\sum_{m=0}^{\infty}(m+1)\sqrt{m+1}\mid e_m\rangle\langle e_{m+1}\mid,\\
NA_{\overline{\qu}}&=&\sum_{m=0}^{\infty}(m+1)\sqrt{m+1}\mid e_{m+1}\rangle\langle e_{m}\mid,\\
A_{\overline{\qu}}N&=&\sum_{m=0}^{\infty}m\sqrt{m+1}\mid e_{m+1}\rangle\langle e_{m}\mid.
\end{eqnarray*}
Thereby we get
$$[N,A_{\qu}] = NA_{\qu}-A_{\qu}N=-A_{\qu},\quad\quad
[N,A_{\overline{\qu}}] =NA_{\overline{\qu}}-A_{\overline{\qu}}N=A_{\overline{\qu}},$$
and we already have $[A_{\qu},A_{\overline{\qu}}]=A_1=\mathbb{I}_{\mathfrak{H}}$.
That is, the right quaternionic operators $A_1, A_{\qu}, A_{\overline{\qu}}$ and $N$ satisfy the Weyl-Heisenberg commutation relations, and therefore these operators together with their commutators form the quaternionic version of the Weyl-Heisenberg algebra.
%%%%%%%%%%%%%%%%%%%%%%%%%%%%%%%%%%%%%%%%%%%%%%%%%%%%%%%%%%%%%%%%%%%%%%%%%%%%%%%%%%%%%%%%%%%%%%%%
\subsection{Upper symbols as differential operators}
Recall the sets $\mathcal{E}=\{|e_m\rangle~~|~~m=0,1,2,\cdots\}$ and $\mathcal{O}=\{\phi_m~~|~~m=0,1,2,\cdots\}$, where $\phi_m$ is as in (\ref{phi}). Let $\mathcal{F}=\{\overline{\phi}_m~~|~~m=0,1,2,\cdots\}$. The sets $\mathcal{E}, \mathcal{O}$ and $\mathcal{F}$ are in 1-1 correspondence to each other and $\mathcal{E}$ is the basis of $\mathfrak{H}$. If one replaces the Hilbert space $\mathfrak{H}$ by the Hilbert space of right regular functions $\mathfrak{H}_{reg}=\overline{span}\mathcal{F}$ or by the Hilbert space of right anti-regular functions $\mathfrak{H}_{a-reg}=\overline{span}\mathcal{O}$, where span means right linear span over $H$, then the actions of the operators $A_{\qu}$ and $A_{\overline{\qu}}$ on the basis sets $\mathcal{O}$ and $\mathcal{F}$ take the form
\begin{eqnarray*}
A_{\qu}\phi_m&=&\sqrt{m}\phi_{m-1},\\
A_{\overline{\qu}}\phi_m&=&\sqrt{m+1}\phi_{m+1},\quad\text{ and }\\
A_{\qu}\overline{\phi}_m&=&\sqrt{m}~~\overline{\phi}_{m-1},\\ A_{\overline{\qu}}\overline{\phi}_m&=&\sqrt{m+1}~~\overline{\phi}_{m+1}.
\end{eqnarray*}
Therefore, on the space $\mathfrak{H}_{a-reg}$ the operator $A_{\qu}=\frac{\partial}{\partial\overline{\qu}}$ and $A_{\overline{\qu}}$ is multiplication by $\overline{\qu}$ , and similarly,  on the space $\mathfrak{H}_{reg}$ the operator $A_{\qu}=\frac{\partial}{\partial{\qu}}$ and $A_{\overline{\qu}}$ is multiplication by ${\qu}$,  where the quaternionic derivatives are right Cullen derivatives (again in
complete analogy with the complex case). For details on Cullen derivatives we refer the reader to \cite{Gen1, Gen2} and for application of Cullen derivatives on quaternionic CS we refer to \cite{Thi1}. Also note that the spaces $\mathfrak{H}_{reg}$ and $\mathfrak{H}_{a-reg}$ are the quaternionic analogue of the Bargmann space of analytic functions and the Bargmann space of anti-analytic functions respectively \cite{Thi1, Al}.

%%%%%%%%%%%%%%%%%%%%%%%%%%%%%%%%%%%%%%%%%%%%%%%%%%%%%%%%%%%%
\subsection{Overlap of the CS and lower symbols}
The overlap of two CS is required to compute the lower symbols. Let $\qu,\pu\in H$, then the overlap of the CS becomes
\begin{eqnarray*}
\langle \gamma_{\qu}\mid \gamma_{\pu}\rangle
&=&\frac{1}{\sqrt{\N(|\qu|)\N(|\pu|)}}\sum_{m=0}^{\infty}\frac{1}{m!}\overline{\qu}^m\pu^m.
\end{eqnarray*}
That is, 
\begin{equation}\label{eqqu1}
\langle \gamma_{\qu}\mid \gamma_{\pu}\rangle=e^{-(|\qu|^2+|\pu|^2)/2}E(\overline{\qu},\pu).
\end{equation}
Now the lower symbol of $A_{f}$ can be computed as 
\begin{eqnarray*}
\tilde{f}&=&\langle \gamma_{\pu}\mid A_{f}\mid\gamma_{\pu}\rangle\\
&=&\int_{H}\langle \gamma_{\pu}\mid\gamma_{\qu}\rangle f(\qu,\overline{\qu})\langle\gamma_{\qu}\mid \gamma_{\pu}\rangle d\varsigma(r, \theta,\phi, \psi)\\
&=&\int_H\Phi(\pu,\qu)d\varsigma(r, \theta,\phi, \psi);
\end{eqnarray*}
where $$\Phi(\pu,\qu)=e^{-(|\qu|^2+|\pu|^2)}E(\overline{\pu},\qu)f(\qu,\overline{\qu})E(\overline{\qu},\pu).$$
%%%%%%%%%%%%%%%%%%%%%%%%%%%%%
\section{Quantization with quaternionic Hermite polynomials}
In this section we present, as illustrative examples, two quantization maps with the quaternionic Hermite polynomials $H_n(\qu)$ and $H_{n,m}(\qu,\oqu)$. In \cite{Thi1} these polynomials and associated CS are discussed in detail. Using the complex counterparts, $H_n(z)$ and $H_{n,m}(z,\oz)$, of these polynomials a non-commutative plane and the complex plane were quantized in \cite{Ga}, \cite{Nic} respectively. By combining the results of \cite{Ga}, \cite{Nic} and \cite{Thi1} we can easily obtain quantization maps and associated results  for the quaternion case. 
\subsection{Quantization map with $H_n(\qu)$}
In \cite{Ga} using the complex Hermite polynomials
\begin{equation}\label{H1}
H_n(z)=n!\sum_{m=0}^{[n/2]}\frac{(-1)^n(2z)^{n-2m}}{m!(n-2m)!};\quad z\in\C 
\end{equation}
the authors have quantized a non-commutative plane. Let $0<s<1$ and $z=x+iy$. The polynomials $H_n(z)$ form an orthogonal set with respect to the measure
$$d\nu_s(z)=d\nu_s(x,y)=exp[-(1-s)x^2-(\frac{1}{s}-1)y^2]dxdy$$
and the normalization factor is
$$b_n(s)=\frac{\pi\sqrt{s}}{1-s}\left(2\frac{1+s}{1-s}\right)^nn!.$$
Take the normalized polynomials $h_{n,s}(z)=b_n(s)^{-1/2}H_n(z)$, then the corresponding reproducing kernel required to the construction of CS is
$$K_s(z,\oz)=\sum_nh_{n,s}(z)\overline{h_{n,s}(z)}
=\frac{1-s^2}{2\pi s}\mbox{exp}\left[\frac{1-s}{2}x^2+\frac{s^2-3s+2}{2s}y^2\right].$$
If $\{|e_m\rangle\}_{m=0}^{\infty}$ is an orthonormal basis for a complex Hilbert space $\HI_{\C}$, then a set of CS associated with $H_n(z)$ can be written as
$$|z,s\rangle=K_s(z,\oz)^{-\frac{1}{2}}\sum_{m=0}^{\infty}\overline{h_{n,s}(z)}|e_m\rangle\in\HI_{\C}.$$
The corresponding quantization map takes the form
$$f(z)\mapsto A_f=\int_{\C}d\nu_s(z)K_s(z,\oz)f(z)|z,s\rangle\langle z,s|.$$
Let $\qu\in H$. Then $\qu$ can be written as
\begin{equation}\label{H2}
\qu=u_{\qu}\mbox{diag}(z,\oz)u_{\qu}^{\dagger}\quad\mbox{with}~~~u_{\qu}\in SU(2),~~z\in\C.
\end{equation}
Let $d\omega(u_{\qu})$ be the normalized Harr measure on $SU(2)$. In \cite{Thi1} using the expression (\ref{H2}) the authors have obtained the quaternionic Hermite polynomials $H_n(\qu)$ which has the same expression as (\ref{H1}) with $z$ replaced by $\qu$. The polynomials $H_n(\qu)$ form an orthogonal set with respect to the measure $d\eta(\qu)=d\nu_s(z)d\omega(u_{\qu})$ with the same normalization constants $b_n(s)$. The corresponding reproducing kernel required to the construction of CS is $K_s(\qu,\oqu)=K_s(z,\oz)\mathbb{I}_2.$ The normalized polynomials are $H_n^s(\qu)=b_n(s)^{-1/2}H_n(\qu).$ Let $\{|\phi_n\rangle\}_{m=0}^{\infty}$ be an orthonormal basis for a $V_H^R$. The set of vectors
$$|\eta_{n,s}\rangle=K_s(\qu,\oqu)^{-\frac{1}{2}}\sum_{m=0}^{\infty}|\phi_m\rangle\overline{H_m^s(\qu)}\in V_H^R$$
forms a set of CS. The corresponding quantization map is
\begin{equation}\label{H3}
A_f=\int_H |\eta_{\qu,s}\rangle f(\qu,\oqu)\langle\eta_{\qu,s}|K_s(\qu,\oqu)d\eta(\qu).
\end{equation}
From (\ref{H3}), using the computations of $A_z$, $A_{\oz}$ and their commutators in \cite{Ga}, one can easily obtain $A_{\qu}, A_{\oqu}$ and their commutators. These computations will be parallel to the proof of the Theorem 4.1 in \cite{Thi1}.
\subsection{Quantization map with $H_{n,m}(\qu, \oqu)$}
The two indexed complex Hermite polynomials are given by
$$h_{n,m}(z,\oz)=(-1)^{n+m}e^{|z|^2}\left(\frac{\partial}{\partial z}\right)^n
\left(\frac{\partial}{\partial \oz}\right)^me^{-|z|^2}.$$
These polynomials form an orthogonal system with respect to the measure $d\nu(z)=\frac{1}{\pi}e^{-|z|^2}d^2z$ and the normalization constant is $n!m!$. In \cite{Thi1} using the same technique of the polynomials $H_n(\qu)$ the authors defined the quaternionic polynomials $H_{n,m}(\qu,\oqu)$. These polynomials form an orthogonal system with respect to the measure $d\tau(\qu)=d\nu(\qu)d\omega(u_{\qu})$ with the normalization constants $n!m!$. The normalized polynomials are $h_{n,m}(\qu,\oqu)=H_{n,m}(\qu,\oqu)/\sqrt{n!m!}$. The reproducing kernel required for the construction of CS is
$$K_n(\qu,\oqu)=\sum_{m=0}^{\infty}h_{n,m}(\qu,\oqu)\overline{h_{n,m}(\qu,\oqu)}=e^{|\qu|^2}\mathbb{I}_2.$$
The convergence and a closed form for this kernel follows from its complex counterpart (see pp.405 of \cite{In} and the proof of lemma 4.2 of \cite{Thi1}). Let $\{|\phi_m\rangle\}_{m=0}^{\infty}$ be an orthonormal basis for a $V_H^R$. A set of CS associated with the polynomials takes the form
\begin{equation}\label{H4}
|\eta_{\qu,n}\rangle=K_n(\qu,\oqu)^{-\frac{1}{2}}\sum_{m=0}^{\infty}|\phi_m\rangle\overline{h_{n,m}(\qu,\oqu)}\in V_H^R.
\end{equation}
The corresponding quantization map is
\begin{equation}\label{H5}
A_f=\int_H|\eta_{\qu,n}\rangle f(\qu,\oqu)\langle\eta_{\qu,n}|K_n(\qu,\oqu)d\tau(\qu).
\end{equation}
In \cite{Nic}, when $n\geq m$ and $m\geq 0$, by writing $h_{n,m}(z,\oz)$ as $h_{n+s,s}(z,\oz)$ the authors have obtained a set of CS similar to the set of CS (\ref{H4}) and used it to quantize the complex plane. Using the results of \cite{Nic} and the quantization map (\ref{H5}) we can obtain the operators $A_{\qu}, A_{\oqu}$ and their commutators. Once again such calculations will be similar to the proof of the Theorem 4.4 in \cite{Thi1}.

%%%%%%%%%%%%%%%%%%%%%%%%%%%%%%%%%%%%
\section{Conclusion}
There were several attempts to define holomorphy for functions of quaternion variables with polynomials as holomorphic functions, for a brief history we refer the reader to \cite{Bu}.  The most promising and recent definition for holomorphy has appeared in \cite{Gen1}, where the authors demonstrated the polynomials of a quaternion variable as a holomorphic (in quaternion language, regular) functions with the aid of the so-called Cullen derivatives.  Reference \cite{Ad} is a treatise on quaternionic quantum mechanics, which describe the theory with real derivatives by considering a function in a quaternion variable $\qu$ as  $f(\qu)=f_0(x_0)+f_1(x_1)i+f_2(x_2)j+f_3(x_3)k; (x_0,x_1,x_2,x_3)\in\mathbb{R}^4$ (a quaternion operator is also defined in the same way), where the non-commutativity still play a part. In this note, we have viewed the quantization of quaternions, through quaternionic CS, without considering their real components. In so doing, we have demonstrated the annihilation and creation operators in terms of Cullen derivatives.\\
 As expected, the operators $A_{{\qu}}$ and $A_{\overline{\qu}}$ act as annihilation and creation operators respectively. The matrix representations of these operators are also similar to the complex case. In complete analogy with the complex harmonic oscillator, the quaternionic version of the harmonic oscillator and Weyl-Heisenberg algebra are obtained.  Even though the non-commutativity of quaternions caused technical difficulties, in most part, the quantization procedure of quaternions followed its complex counterpart (except the momentum operator).\\
 A proper definition of quaternionic derivative (the so-called Cullen derivative) \cite{Bu}, quaternionic holomorphy, quaternionic Fock space \cite{Al} and the so-called S-spectrum of a quaternionic operator \cite{Fab, Fab1, Fab2} have been developed very recently (one may also consult the review article \cite{Gi} and the references therein). In this respect, understanding a quaternionic Hamiltonian, as a quaternionic Hermitian operator, is in its early stage.\\
However, some progress has been made. For example, in \cite{Qu} the quaternionic canonical CS were interpreted as CS of a Jaynes-Cummings model when the CS were considered as vectors in the state Hilbert space of the Jaynes-Cummings operator. In \cite{Qu} the authors have also studied some physical properties of the model using these CS. The state Hilbert space of the Jaynes-Cummings model consists vectors of the form
$\left(\begin{array}{c}\psi_n^+\\\psi_n^-\end{array}\right).$
If $\{|\phi_n\rangle\}_{n=0}^{\infty}$ is a basis for a complex Hilbert space of functions (for example a $L^2$ space, or some space where the conjugate make sense) then it is not difficult to see that
$$\left(\begin{array}{cc}
|\phi_n\rangle&0\\
0&\overline{|\phi_n\rangle}\end{array}\right);\quad n=0,1,2,\cdots$$
is a basis for some function space $V_H^R$. Hence, the state Hilbert space of the Jaynes-Cummings model can be identified with $V_H^R$ (with a Hilbert space isomorphism). In this regard, the quantization presented in this note may be applied to Jaynes-Cummings models.\\
There is also a purpose to emphasize quaternionic violations of complex quantum mechanics which could be tested in laboratory experiments. The recent progress and interest in looking for quantitative and qualitative differences between complex and quaternionic quantum mechanics may also have an impact in our interest. For these lines of arguments see \cite{Leo} and the many references therein. \\
The quantization procedure plays an important role in complex quantum mechanics.  The quantization presented in this manuscript is likely to play a role in quaternionic quantum mechanics.
%%%%%%%%%%%%%%%%%%%%%%%%%%%%%%%%%%%%%%%%%%%%%%%%%%%%%%%%%%%%%%%%%%%%%%%%%%%%%%%%%%%%%%%%%%%%%%%%%%%%%%%%%%%%%%%%%%%%%%%%%%%%%%%%%%%%%%%%%%%%%%%%%%%%%%%%%%%%%%%%%%%%%

%\end{multicols}
%%%%%%%%%%%%%%%%%%%%%%%%%%%%%%%%%%%%%%%%%%%%%%%%%%%%%%%%%%%%%%%%%%%%%%%%%%%%%%%%%%%%%%%%%%%%%%%%%%%%%%%%%%%%%%%%%%%%%%%%%%%%%%%%%%%%%%%%%%%%%%%%%%%%%%%%%%%%%%%%%%%%

\begin{thebibliography}{XXXX}

\bibitem{Ad} Adler, S.L., {\em Quaternionic quantum mechanics and Quantum fields}, Oxford University Press, New York, 1995.

\bibitem{Ali} Ali, S.T., Antoine, J-P., Gazeau, J-P., {\em Coherent States, Wavelets and Their Generalizations}, Springer,  New York, 2000.

\bibitem{AE} Ali, S.T., Englis, M., {\em Quantization method: a guide for Physicist and analysts}, Rev. Math. Phys., {\bf 17} (2005), 391-490.

\bibitem{Al} Alpay, D., Colombo, F., Sabadini, I., Salomon, G., {\em The Fock space in the slice hyperholomorphic setting}, Hypercomplex Analysis: New perspective and applications, Trends in Mathematics, Birkh\"user, Basel (2014), 43-59.

\bibitem{AGH} Aremua, I., Gazeau, J-P., Hounkonnou, M.N., {\em Action-angle coherent states for quantum systems with cylindrical phase space}, J. Phys. A. {\bf 45} (2012), 335302.

\bibitem{Bu}Buchmann, A., {\em A brief history of quaternions and the theory of holomorphic functions of quaternion variable}, arXiv:1111.6088[Math.HO].

\bibitem{Fab}Colombo, F., Sabadini, I., \textit{On some properties of the quaternionic functional calculus}, J. Geom. Anal., {\bf 19} (2009), 601-627.

\bibitem{Fab1}Colombo, F., Sabadini, I., \textit{On the  formulations of the quaternionic functional calculus}, J. Geom. Phys., {\bf 60} (2010), 1490-1508.

\bibitem{Fab2} Colombo, F., Sabadini, I., Struppa, D.C., {\em Noncommutative functional calculus}, {\bf 289}, Progress in Mathematics, Birkhauser, 2011.

\bibitem{Nic} Cotfas, N., Gazeau, J-P., Gorska, K., {\em Complex and real Hermite polynomials and related quantizations}, J. Phys.A: Math. Theor. {\bf 43} (2010), 305304.

\bibitem{Leo} De Leo, S., Giardino, S., {\em Dirac solutions for quaternionic potentials}, J. Math. Phys. {\bf 55} (2014), 022301.

\bibitem{Eb} Ebbinghaus, H.D., Hermes, H., Hirzebruch, F., Koecher, M., Mainzer, K., Neukirch, J., Prestel,
A., Remmert, R., {\em Numbers},(3rd Edition), Springer, New York (1995).

\bibitem{In} Intissar, A., Intissar, A., {\em Spectral properties of the Cauchy transform on $L_2(\C, e^{-|z|^2}\lambda(z))$}, J. Math. Anal. Appl. {\bf 313} (2006), 400-418.

\bibitem{Gaz} Gazeau, J-P., {\em Coherent states in quantum physics}, Wiley-VCH,  Berlin (2009).

\bibitem{Ga} Gazeau, J-P., Szafraniec, F.H., {\em Holomorphic Hermite polynomials and non-commutative plane}, J. Phys. A: Math. Theor. {\bf 44} (2011), 495201.
\bibitem{Gi}Ghiloni, R., Moretti, W. and Perotti, A., {\em Continuous slice functional calculus in quaternionic
Hilbert spaces}, Rev. Math. Phys. {\bf 25} (2013), 1350006..

\bibitem{Gen1} Gentili, G.,Struppa, D.C., {\em A new theory of regular functions of a quaternionic variable}, Adv. Math. {\bf 216} (2007), 279-301.

\bibitem{Gen2} Gentili, G., Stoppato, C., {\em  Power series and analyticity over the quaternions}, Math. Ann. {\bf 352} (2012), 113-131.

\bibitem{Gu} G\"ulebeck, K., Habetha, K., Spr\"obig, W., {\em Holomorphic functions in the plane and n-dimensional spaces}, Birkh\"auser Verlag, Basel (2008).

\bibitem{Mo} Mouayn, Z., {\em Coherent states quantization for generalized Bargmann spaces with formulae for their attached Berezin transform in terms of the Laplacian on $\mathbb{C}^{n}$}, J. Fourier. Anal. Appl. {\bf 18}, (2012), 609-625.

\bibitem{Per} Perelomov, A., {\em Generalized coherent states and their applications}, Springer-Verlag, Berlin (1986).


\bibitem{Thi1}Thirulogasanthar, K., Twareque Ali, S., {\em Regular subspaces of a quaternionic Hilbert
space from quaternionic Hermite polynomials and associated coherent states}, J. Math. Phys., {\bf 54} (2013), 013506.

\bibitem{Thi2} Thirulogasanthar, K., Honnouvo, G., Krzyzak, A., {\em Coherent states and Hermite polynomials on Quaternionic Hilbert spaces}, J. Phys.A: Math. Theor. {\bf 43} (2010), 385205.

\bibitem{Qu} Thirulogasanthar, K., Krzy\.zak, A, Katatbeh, Q.D., {\em Quaternionic vector coherent states and the supersymmetric Harmonic oscillator}, Theor. Math. Phys. {\bf 149} (2006), 1366-1381.

\bibitem{Ale} Torgasev, A., {\em Dual space of a quaternion Hilbert space}, Ser. Math. Inform. 14 (1999), 71-77.

\bibitem{Vis} Viswanath, K., {\em Normal operators on quaternionic Hilbert spaces}, Trans. Am. Math. Soc. {\bf 162} (1971), 337�350.

\bibitem{Za} Zhang, F., {\em Quaternions and Matrices of Quaternions}, Linear Algebra and its Applications, {\bf 251} (1997), 21-57.

\end{thebibliography}
\end{document}